\documentclass[aps,prd,twocolumn,nofootinbib,superscriptaddress]{revtex4}
\pdfoutput=1
\usepackage{amsmath,amssymb}
\usepackage{bm,bbm}
\usepackage{graphicx, graphics, color}
\usepackage{slashed}

\newcommand{\fref}[1]{Fig.~\ref{f.#1}}

\newcommand{\eref}[1]{Eq.~(\ref{e.#1})}

\newcommand{\sref}[1]{Section~\ref{s.#1}}

\newcommand{\tref}[1]{Table~\ref{t.#1}}
\newcommand{\beq}{\begin{eqnarray}}
\newcommand{\eeq}{\end{eqnarray}}
\newcommand{\beqs}{\begin{eqnarray*}}
\newcommand{\eeqs}{\end{eqnarray*}}

\newcommand{\co}{{\mathcal O}}

\newcommand{\GeV}{\>\text{GeV}}

\newcommand{\SU}{\mathrm{SU}}
\newcommand{\U}{\mathrm{U}}

\begin{document}

\title{Prospect for measuring the CP phase in the $h\tau\tau$ coupling at the LHC}

\author{Andrew Askew}
\affiliation{Department of Physics, Florida State University, Tallahassee, FL 32306, USA}

\author{Prerit Jaiswal}
\affiliation{Department of Physics, Syracuse University, Syracuse, NY 13244, USA}
\affiliation{Department of Physics, Florida State University, Tallahassee, FL 32306, USA}

\author{Takemichi Okui}
\affiliation{Department of Physics, Florida State University, Tallahassee, FL 32306, USA}

\author{Harrison B. Prosper}
\affiliation{Department of Physics, Florida State University, Tallahassee, FL 32306, USA}

\author{Nobuo Sato}
\affiliation{Jefferson Lab, Newport News, Virginia 23606, USA}
\affiliation{Department of Physics, Florida State University, Tallahassee, FL 32306, USA}

\begin{abstract}
The search for a new source of CP violation is one of the most
important endeavors in particle physics. A particularly interesting
way to perform this search is to probe the CP phase in the $h\tau\tau$
coupling, as the phase is currently completely unconstrained by all
existing data.  Recently, a novel variable $\Theta$ was proposed for
measuring the CP phase in the $h\tau\tau$ coupling through the
$\tau^\pm \to \pi^\pm \pi^0 \nu$ decay mode.   We examine two crucial
questions that the real LHC detectors must face, namely, the issue of
neutrino reconstruction and the effects of finite detector resolution.  
For the former, we find strong evidence that the
collinear approximation is the best for the $\Theta$ variable.  For
the latter, we find that the angular resolution is actually not an
issue even though the reconstruction of $\Theta$ requires resolving
the highly collimated $\pi^\pm$'s and $\pi^0$'s from the $\tau$
decays. Instead, we find that it is the missing transverse energy
resolution that significantly limits the LHC reach for measuring the
CP phase via $\Theta$. With the current missing energy resolution, we find that with
$\sim 1000\,\textrm{fb}^{-1}$ the CP phase
hypotheses $\Delta = 0^\circ$ (the standard model value) and $\Delta = 90^\circ$
can be distinguished, at most, at the 95\% confidence level.
\end{abstract}

\maketitle

\section{Introduction}
\label{s.intro}
The Standard Model (SM) of particle physics has been extremely
successful in explaining all microscopic phenomena observed down to
distances of order $\sim 1\textrm{\,TeV}^{-1}$. However, the  model
has two notable under-explored corners: the neutrino and Higgs
sectors.  Having discovered the Higgs boson the next major goal of
the Large Hadron Collider (LHC) is to scrutinize the properties of the
Higgs boson and search for deviations from the SM predictions. 



A particularly interesting question is whether CP is a good symmetry
of the Higgs sector.  Since all Higgs boson couplings are predicted to
be CP even in the SM, any observation of a CP-odd or CP-violating
component will constitute unambiguous evidence of physics beyond the
SM\@.  CP-violating Higgs interactions may be motivated by the fact
that baryogenesis requires the existence of a new source of CP
violation and the possibility that such a primordial source of new CP
violation may leave traces in the couplings of the Higgs boson.

Recently, a novel variable $\Theta$ for measuring the CP phase in the
$h$-$\tau$-$\tau$ coupling was proposed in Ref.~\cite{Harnik:2013aja}.
The distribution of $\Theta$ was shown to have the simple form
$\displaystyle{c - a \cos(\Theta - 2\Delta)}$, where $\Delta$ is the
phase of the $h$-$\tau$-$\tau$ Yukawa coupling, with $\Delta =
0^\circ$ and $90^\circ$ corresponding to the CP-even and -odd limits,
respectively, while $a$ and $c$ are independent of $\Delta$ and
$\Theta$.  What distinguishes the $h$-$\tau$-$\tau$ Yukawa coupling
from other couplings is that its \emph{phase} (i.e.~the CP angle
$\Delta$) is currently completely unconstrained by
data~\cite{Brod:2013cka}, even though its \emph{magnitude} (i.e.~the
rate of $\tau^+$-$\tau^-$ production) is roughly consistent with
the SM prediction~\cite{Chatrchyan:2014nva, ATLAS:2014htt}.  Field
theoretically, as discussed in Ref.~\cite{Harnik:2013aja}, an $\co(1)$
$\Delta$ in the $h$-$\tau$-$\tau$ coupling can easily coexist with
other Higgs boson couplings whose CP phases are known to be small,
such as $h$-$t$-$\bar{t}$ and $h$-$b$-$\bar{b}$ ($\Delta_t \lesssim
0.01$ and $\Delta_b \lesssim 0.1$, respectively~\cite{Brod:2013cka}),
and $h$-$Z$-$Z$ (for which $\Delta_Z = 90^\circ$ is
disfavored~\cite{Chatrchyan:2013mxa,Aad:2014eva,D0:2013rca}). In
Ref.~\cite{Harnik:2013aja} using parton-level analyses  it
is estimated that the $h$-$\tau$-$\tau$ CP phase, $\Delta$, may be
measured from the $\Theta$ distribution to an accuracy of $\sim
11^\circ$ at the 14-TeV LHC with $\sim 3000\textrm{\,fb}^{-1}$ of
data, or the $\Delta = 0^\circ$ and $\Delta = 90^\circ$ cases may be
distinguished at $5\sigma$ level with $\sim 1000\textrm{\,fb}^{-1}$.
However, these conclusions are for an ideal  detector. In this paper,
we examine the impact of realistic detector effects on the measurement
of $\Delta$ via $\Theta$.

There are various ways that detector effects can affect those
estimates.  Most importantly, the reconstruction of the $\Theta$
variable requires resolving the charged and neutral pions in the
$\tau^\pm \to \rho^\pm \nu \to \pi^\pm \pi^0 \nu$ decay mode, but
these $\tau^\pm$ are highly boosted as they come from higgs decays.
One therefore expects the two photons from each $\pi^0$ decay to hit
the same tower of the electromagnetic calorimeter (ECAL), thereby
making the standard identification of a $\pi^0$ using photons useless.
Moreover, one expects the same ECAL tower to be hit by the $\pi^\pm$,
raising the question of how much of the ECAL tower's activity should be
attributed to the $\pi^0$.

Another obvious issue is the reconstruction of the neutrinos.  In
Ref.~\cite{Harnik:2013aja}, the so-called collinear
approximation~\cite{Ellis:1987xu} is adopted, where the neutrino from
each $\tau$ decay is taken to be exactly collinear with its parent
$\tau$. As shown in Ref.~\cite{Harnik:2013aja}, the collinear
approximation results in a 60--70\% reduction of the amplitude of the
$\cos(\Theta - 2\Delta)$ curve.  So, one might expect that here lies a
significant opportunity for improvement.  In this paper, however, we
present a theoretical calculation that strongly suggests that it is
extremely difficult, if not impossible, to outperform the collinear
approximation for the purpose of $\Theta$ reconstruction. 

Adopting the collinear approximation, we then focus on the
aforementioned issues of pion identifications in the presence of
realistic detector effects, and propose a realistic algorithm to
identify ``$\tau$ candidates'' specifically suited for constructing
$\Theta$.  Surprisingly, we find that the amplitude of the
$\cos(\Theta - 2\Delta)$ curve of the signal is virtually unchanged by
the finite angular resolution of the LHC detectors.  Instead, we find
that the measurement of $\Delta$ through the $\Theta$ distribution is
significantly hampered by the contamination by the $Z \to \tau\tau$
background. This degradation arises due to resolution effects in
reconstructing the $\tau\bar{\tau}$ invariant mass as a consequence of
large uncertainties in the measurement of the missing transverse
energy. 
Fortunately, this is an area where future improvement
can be hoped for, unlike the angular resolution.  We leave the
important question of pileup effects for future work. 

While we focus on the measurement of $\Delta$ through the $\Theta$
variable, similar but different acoplanarity angles have been studied
both for $e^+ e^-$ colliders~\cite{CPtau:ee1,CPtau:ee2,CPtau:ee3,
CPtau:ee4, CPtau:ee5,CPtau:ee6} as well as for the LHC~\cite{CPtau:Berge1,
CPtau:Berge2, CPtau:Berge3,CPtau:Berge4, CPtau:Berge5, CPtau:Berge6}.
For the $\tau^\pm \to \pi^\pm \pi^0 \nu$ decay mode, which is the most
dominant decay mode of the $\tau$, the acoplanarity angles used in those
studies are not as effective as the $\Theta$ variable at $e^+ e^-$
colliders in which the neutrino momenta can be reconstructed up to a
two-fold ambiguity~\cite{Harnik:2013aja}.  An advantage of the LHC
studies mentioned above is that they can be applied to all 1-prong
$\tau$ decays including leptonic decays.  However, realistic detector
effects are not considered in those studies.  Ref.~\cite{CPtau:H2j}
performs a realistic study for the LHC but assuming universal CP
violation for all SM fermions, unlike our study in which we assume
$\Delta$ is non-zero only for the $\tau$.  Finally, the use of the $\tau$
impact parameter associated with a displaced $\tau$ vertex was studied
in Ref.~\cite{CPtau:eeDisplaced} in the $e^+ e^-$ context.  Our $\Theta$
variable does not depend on the measurement of the $\tau$ impact
parameter.

The paper is organized as follows. In \sref{theory}, we review the
$\Theta$ variable and rewrite it in a simple form by expanding it in
terms of a small parameter $\lambda$ that characterizes the degree of
collinearity of the decay products of the taus, with $\lambda \to 0$
in the exactly collinear limit.  We use this expansion to argue that
the collinear approximation appears the best we can do for the purpose
of reconstructing $\Theta$.  \sref{parton} describes the parton level
simulation of the signal and background processes.  \sref{simulation}
describes the detector simulation and selection of the parametrized
detector quantities corresponding to the Higgs boson decay products.
\sref{results} evaluates the results of the simulation and sensitivity
as a function of integrated luminosity, and we conclude in
\sref{conc}.

\section{Theory}
\label{s.theory}

\subsection{The CP phase $\Delta$ and the variable $\Theta$}
Following Ref.~\cite{Harnik:2013aja}, we define the CP phase $\Delta$ by
\begin{equation}
\mathcal{L}_\text{$\tau$-Yukawa}
= - \frac{y_\tau}{\sqrt2} h \bar{\tau} (\cos\Delta + \mathrm{i} \gamma_5 \sin\Delta) \tau 
\,,\label{e.Yukawa}  
\end{equation}
where the scalar field $h$ and Dirac spinor $\tau$ describe the Higgs
boson and $\tau$ lepton, respectively. In order to maintain the SM
decay rate for $h \to \tau\tau$, we fix the $h$-$\tau$-$\tau$  Yukawa
coupling, $y_\tau$, to its SM value $y_\tau^\text{SM} = m_\tau / v$
with $v \simeq 174\GeV$, but keep the CP phase $\Delta$ as a free
parameter to be measured. The SM prediction is $\Delta = 0$.%
\footnote{ For a fully $\SU(2) \times \U(1)$ gauge-invariant form of
the interaction~(\ref{e.Yukawa}) as well as an example of
renormalizable realizations of it, see Ref.~\cite{Harnik:2013aja}.}

We are interested in the decay chain
\begin{equation}
\begin{split} 
h 
&\to \tau^+ \tau^- \\
&\to \rho^+ \bar{\nu} \, \rho^- \nu \\
&\to \pi^+ \pi^0 \bar{\nu} \, \pi^- \pi^0 \nu \,.
\end{split}
\end{equation}
As calculated in Ref.~\cite{Harnik:2013aja}, the squared matrix
element for this decay has the form
\begin{equation}
|\mathcal{M}_{h \to \pi\pi \bar{\nu} \, \pi \pi \nu}|^2
\propto
c - a \cos(\Theta - 2\Delta)
\,,
\label{e.amp2}
\end{equation}
where $a$ and $c$ are both positive and depend neither on $\Theta$ nor
$\Delta$.  The angular variable $\Theta$ is defined through
\begin{equation}
\begin{split}
\cos\Theta 
&\propto
(k_1 \!\cdot\! p_{\tau 2}) \, (k_2 \!\cdot\! p_{\tau 1}) 
- (p_{\tau 1} \!\cdot\! p_{\tau 2}) \, (k_1 \!\cdot\! k_2)
\,,\\
\sin\Theta 
&\propto
\epsilon_{\mu \nu \rho \sigma} \,
k_1^{\mu} \, p_{\tau 1}^\nu \, k_2^\rho \, p_{\tau 2}^\sigma 
\,,
\end{split}
\label{e.Theta}
\end{equation}    
with a common proportionality factor.  The subscripts $1$ and $2$
refer to the $\tau^+$ and $\tau^-$ branches of the decay, respectively
(e.g., $p_{\tau 1}^\mu = p_{\tau^+}^\mu$ and $p_{\tau 2}^\mu =
p_{\tau^-}^\mu$).  The 4-momenta $k_i^\mu$ ($i=1$, $2$) are defined by
\begin{equation}
k_i^\mu 
\equiv 
y_i q_i^\mu + r p_{\nu i}^\mu,
\label{e.k}
\end{equation}
with 
\begin{equation}
q_i^\mu \equiv p_{\pi^\pm i}^\mu - p_{\pi^0 i}^\mu 
\,,\label{e.q}
\end{equation}
and
\begin{equation}
y_i \equiv \frac{2 q_i \!\cdot\! p_{\tau i}}{m_\tau^2 + m_\rho^2} 
\,,\quad  
r \equiv \frac{m_\rho^2 - 4 m_\pi^2}{m_\tau^2 + m_\rho^2} \approx 0.14
\,.
\end{equation}
%

\subsection{The boosted $\tau^\pm$ limit}
The fact that the taus from Higgs boson decay are highly boosted can
be exploited to obtain an even simpler expression for $\Theta$.  We
define two lightlike 4-vectors $n_1^\mu$ and $n_2^\mu$ whose spatial
components, $\vec{n}_1$ and $\vec{n}_2$, are unit vectors taken along
the 3-momenta of the $\rho^\pm$ from the $\tau$ decays:
\begin{equation}
n_1 \equiv (1, \hat{n}_{\rho 1})
\,,\quad
n_2 \equiv (1, \hat{n}_{\rho 2})
\,.\label{e.n1_and_n2}
\end{equation} 
We also define
\begin{equation}
\bar{n}_1 \equiv (1, -\hat{n}_{\rho 1})
\,,\quad
\bar{n}_2 \equiv (1, -\hat{n}_{\rho 2})
\,.
\end{equation} 
Then, an arbitrary 4-momentum $a^\mu$ 
can be expanded in terms of $n_1$ and $\bar{n}_1$ as
\begin{equation}
\begin{split}
a^\mu 
&= (a \cdot \bar{n}_1) \, \frac{n_1^\mu}{2} 
   + (a \cdot n_1) \frac{\bar{n}_1^\mu}{2} + a^{\perp \mu}  \\
&\equiv a^+ \frac{n_1^\mu}{2} + a^- \frac{\bar{n}_1^\mu}{2} + a^{\perp \mu}  \,, 
\end{split}
\end{equation}
where $a^\perp$ by definition satisfies $\displaystyle{a^\perp \!\cdot
n_1} = \displaystyle{a^\perp \!\cdot \bar{n}_1} = 0$.  Alternatively,
$a^\mu$ can be expanded in a similar manner in terms of $n_2$ and
$\bar{n}_2$.  To avoid notational clutter, we do not make explicit
whether $a^\pm$ and $a^\perp$ are defined in the $n_1$-$\bar{n}_1$
basis or the $n_2$-$\bar{n}_2$ basis,  but it should be clear that the
momenta of all particles originating from the $\tau^+$ will be
expanded in the $n_1$-$\bar{n}_1$ basis, while those from $\tau^-$
will be in $n_2$-$\bar{n}_2$.
  
Next, we introduce a small expansion parameter $\lambda \ll 1$ to
quantify the degree of collinearity between the $\tau^\pm$ and the
respective $\rho^\pm$.  First, by construction, ``collinear'' means
that $p^\mu$ is dominated by the $+$ component, $p^+$, while the
remaining components $p^\perp$ and $p^-$ are small.  Therefore, by
definition, we assign the $\lambda$ scaling law $p^\perp \sim \lambda
p^+$ to the $\perp$ component.  This implies that numerically we have
$\lambda \sim m_\tau / m_h \sim 10^{-2}$.  Then, in the
ultra-relativistic limit, we have $0 \simeq \displaystyle{p \cdot p} =
\displaystyle{p^+ p^- + p^\perp \!\cdot p^\perp} \sim
\displaystyle{p^+ p^- + \mathcal{O}(\lambda^2 p^+ p^+)}$, so we see
that $p^- \sim \lambda^2 p^+$.  To summarize, we characterize the
collinearity of $p$ as
\begin{equation}
(p^+, p^-, p^\perp) \sim (1, \lambda^2, \lambda) \,p^+ 
\label{e.lambda_scaling}\,.
\end{equation}
%

\subsection{The reconstruction of neutrinos}
Since neutrinos are invisible at the LHC, we must devise a method to
assign momenta to the neutrinos from the $\tau^\pm$ decays.  However,
the choice of method is a delicate issue for the reconstruction of the
$\Theta$ variable.  Performing the $\lambda$ expansion in the
expressions~(\ref{e.Theta}), we see that all $\mathcal{O}(\lambda^0)$
and $\mathcal{O}(\lambda^1)$ terms exactly cancel, so we are left with
\begin{equation}
\begin{split}
(k_1 \!\cdot\! p_{\tau 2}) \, (k_2 \!\cdot\! p_{\tau 1}) 
- (p_{\tau 1} \!\cdot\! p_{\tau 2}) \, (k_1 \!\cdot\! k_2)
&\sim \mathcal{O}(\lambda^2)
\,,\\
\epsilon_{\mu \nu \rho \sigma} \,
k_1^{\mu} \, p_{\tau 1}^\nu \, k_2^\rho \, p_{\tau 2}^\sigma 
&\sim \mathcal{O}(\lambda^2)
\,.
\end{split}
\label{e.cancellations}
\end{equation}    
Therefore, for the purpose of reconstructing $\Theta$, values must be
carefully assigned to the invisible neutrino momenta so as not to
disturb these delicate cancellations.  In particular, the $\lambda$
scaling law~(\ref{e.lambda_scaling}) must be respected.

A simple method with this property is provided by the \emph{collinear
approximation}~\cite{Ellis:1987xu}, where we set
\begin{equation}
p_{\tau i}^{\perp\mu} = p_{\tau i}^- = 0
\,,\quad
p_{\nu i}^{\perp\mu} = p_{\nu i}^- = 0\,,
\end{equation}
for each $i$.  In other words, the collinear approximation applies the
$\lambda \to 0$ limit to the $\tau$ and $\nu$ momenta.  Then, to
$\mathcal{O}(\lambda^2)$, the expressions~(\ref{e.Theta}) become 
\begin{equation}
\begin{split}
\cos\Theta 
&\propto
(k_1^{\perp} \!\cdot n_2) \, (k_2^{\perp} \!\cdot n_1) 
- (n_1 \!\cdot n_2) \, (k_1^\perp \!\cdot\! k_2^\perp)
\,,\\
\sin\Theta 
&\propto
\epsilon_{\mu \nu \rho \sigma} \,
k_1^{\perp\mu} n_1^\nu \, k_2^{\perp\rho} n_2^\sigma 
\,,
\end{split}
\end{equation}    
again with a common proportionality factor.  Finally, noticing that
the definitions of $k_i^\mu$ in~(\ref{e.k}) give $k_i^{\perp\mu} = y_i
q_i^{\perp\mu}$ in the collinear approximation,  we arrive at the
following very simple expressions that determine $\Theta$:
\begin{equation}
\begin{split}
\cos\Theta 
&\propto
(q_1^{\perp} \!\cdot n_2) \, (q_2^{\perp} \!\cdot n_1) 
- (n_1 \!\cdot n_2) \, (q_1^\perp \!\cdot q_2^\perp)
\,,\\
\sin\Theta 
&\propto
\epsilon_{\mu \nu \rho \sigma} \,
q_1^{\perp\mu} n_1^\nu \, q_2^{\perp\rho} n_2^\sigma 
\,,
\end{split}
\label{e.Theta:coll_approx}
\end{equation}    
with a common proportionality factor.  The 4-vectors $q_i^\mu$,
$n_1^\mu$ and $n_2^\mu$ are defined in~(\ref{e.q})
and~(\ref{e.n1_and_n2}), respectively.  In the rest of the paper, we
use the expressions~(\ref{e.Theta:coll_approx}) to compute $\Theta$
unless noted otherwise.

One might think we should be able to do better than the collinear
approximation by using statistical correlations of the neutrino
momenta with other, visible, momenta in the process.  However, the
delicate cancellation shown in (\ref{e.cancellations}) implies that
the correlations must get the neutrino momenta correctly at the level
of $\mathcal{O}(\lambda^2) \sim 10^{-4}$ in order to outperform the
collinear approximation. Although we could not prove that this is
impossible, it certainly suggests that it is extremely difficult. 


\section{Parton-level Simulation}
\label{s.parton}
In this section, we outline our generation of the
parton level samples used in the detector simulation.  We generated
tree level event samples with a combination of {\tt
MadGraph5}~\cite{Alwall:2014hca} and {\tt MCFM} as explained below for
the background ($Z+j$) and the signal ($h+j$) with $p_\text{Tmin}=70$
GeV for the jet. 
The associated jet facilitates triggering and 
provides sufficient $\tau^+\tau^-$ transverse momentum for the
collinear approximation to hold.  Other associated
production modes could be used, notably $h+Z$ and $h+W$, but they 
have lower cross sections than $h+j$, though potentially better signal to noise
that could be exploited in a further study.
Multijet events are a potential background; however, this 
background is found to be small once tau selection criteria are imposed.
Therefore, consideration of $Z+j$ alone is sufficient for our purposes.
  
The jet-$p_\mathrm{T}$ cut was imposed at the generator level in anticipation of the high-$p_\mathrm{T}$ jet
requirement we impose in our analysis to ensure that the
event will pass the trigger as well as to avoid kinematic
configurations where the $\tau^+\tau^-$ is produced back-to-back in
the transverse plane and the collinear approximation for the
neutrino momenta fails.

The model files were generated using {\tt FeynRules}
package~\cite{Degrande:2011ua} by modifying the SM template provided
by {\tt FeynRules} to include the CP violating coupling in the
$h$-$\tau$-$\tau$ vertex as well as the effective vertices for the
decay of $\tau\rightarrow \rho+\nu$  and $\rho\rightarrow\pi+\pi^{0}$
given by
\begin{equation}
\begin{split}
\mathcal{L}_\text{eff}
=
-\rho_{\mu}^{*} \, \bar{\nu} \gamma^{\mu} P_{\text{\tiny L}} \tau 
- \rho^{\mu}
\!\left (\pi^0 \partial_{\mu}\pi^* - \pi^* \partial_{\mu}\pi^0 \right)\! 
+\text{c.c.} \,,
\end{split}
\label{e.Leff}
\end{equation}
where $P_{\text{\tiny L}} \equiv (1 - \gamma_5) / 2$ is the projection
operator onto left-handed chirality. We implement the Higgs boson
production from $gg$ fusion using the effective field theory (EFT) (in which
the top quark is integrated out) available from the {\tt FeynRules}
repository.  In principle, such an effective vertex is applicable only
when the energy scale of the process lies well below the mass of the
top quark.  However, using the {\tt MCFM}
program~\cite{Campbell:2011bn}, we have verified that the Higgs boson
$p_\mathrm{T}$ distribution from the EFT in the region of interest
($70<p_\mathrm{T}<200$ GeV) agrees at the percent level with the full
1-loop QCD calculation.  Finally, note that all the coupling constants in
\eref{Leff} are set to unity.  To get the physical cross section, we
use the cross sections from {\tt MCFM} re-weighted by
$\tau\rightarrow\rho+\nu$ and $\rho\rightarrow \pi+\pi^0$ branching
fractions. The CTEQ6l1 PDF set was used throughout.

Parton showering and hadronization effects are included by processing
the {\tt Madgraph5} samples using the {\tt Pythia8}
program~\cite{Sjostrand:2007gs}.  In principle, higher order
corrections can be implemented by including events with higher jet
multiplicity from tree level matrix elements and performing the
corresponding jet merging with {\tt Pythia}'s parton showering.
However, we have verified that the inclusion of these corrections do
not affect the Higgs boson $p_\mathrm{T}$ distribution and they differ
at most by a few percent in the region of interest.

\section{Detector Simulation}
\label{s.simulation}
In order to incorporate detector effects into the measurement of the
$\Theta$ angle, we use the Snowmass detector model~\cite{Anderson:2013kxz},
implemented in {\tt Delphes~3}~\cite{deFavereau:2013fsa}.  
In {\tt Delphes}, energy from particles deposited
within the calorimeter is partitioned between the electromagnetic and
hadronic sections according to a fixed ratio. This ratio is unity for
the fraction of energy from a charged pion deposited in the hadronic
calorimeter, which fails to account for energy depositions from
hadronic showers that begin before the hadronic calorimeter.  We
account for this effect, including potential systematic effects due to
overlap between the energy deposition of the neutral pion and the
charged pion,  by distributing the energy of charged pions in both
calorimeter sections according to a probability distribution that
models the one given in Ref.~\cite{Abdullin:2009zz} for 30\,GeV pions.
85\% percent of the time the particle is presumed to deposit energy
via ionization only, and the energy is entirely deposited within the
hadronic calorimeter.  For the remaining 15\% of events, the fraction
of the particle energy deposited in the electromagnetic calorimeter is
drawn from a Gaussian distribution of mean 0.4 and width 0.3
(selecting only physical values).  The rest of the energy is assigned
to the hadronic calorimeter.  The Gaussian models the probability with
which the pion interacts early, in which case a significant fraction
of its energy will be deposited in the ECAL.

Tau identification in {\tt Delphes} is abstracted to an efficiency,
which is insufficient  for our purposes.  Therefore, we develop an
algorithm to identify taus explicitly that assigns the charged pion
and the neutral pion to a track and a calorimeter tower, respectively.
We select jets reconstructed with the anti-$k_\mathrm{T}$
algorithm~\cite{Cacciari:2008gp} with a distance parameter of 0.5,
which are reconstructed in {\tt Delphes} using {\tt
FastJet}~\cite{Cacciari:2011ma}.  At least three jets are required in
each event with $p_\mathrm{T} > 30$\,GeV within $|\eta|<2.5$ (the
tracking limit of the detector), of which the lead jet must have
$p_\mathrm{T} > 100$\,GeV (well above the parton level requirement of
70\,GeV).  All jets with $p_\mathrm{T} >$30\,GeV are examined as
potential $\tau$-candidates.  The lead track within $\Delta R<0.2$ of
the jet axis is selected, and the sum of the transverse momenta of the
remaining tracks, within $\Delta R<0.4$, is required to be less than
7\,GeV in order to exclude jets that are unlikely to be taus. These
selection requirements emulate the more sophisticated
selection of taus in the LHC experiments, and are treated as
equivalent to the {\tt Delphes} efficiency of 65\%.

We make further selections on the individual elements within the tau
candidate in  order to ensure that the decay products have been
measured with sufficient precision for the calculation of $\Theta$.
The lead track is required to have $p_\mathrm{T} >5$\,GeV, or this
candidate is rejected. We then proceed to examine calorimeter towers
within $\Delta R<0.2$ of the jet axis.  We select the tower with the
largest transverse energy in the electromagnetic portion of the tower.
The tower in question must also have an electromagnetic fraction
(EMF) of 0.9, that is, at least 90\% of its total energy within the
ECAL.  If this tower has $p_\mathrm{T}>5$\,GeV, is within $\Delta
R<0.2$ of the selected track, and the sum of the selected tower
transverse energy and the track momentum is greater than 20\,GeV, then
this jet becomes a $\tau$-candidate suitable for $\Theta$
reconstruction, and the charged pion is assumed to be the selected
track with the neutral pion the selected tower.

Finally, 
the electromagnetic portion of the actual ATLAS and CMS detectors have finer
granularity than a single tower, which is not properly accounted for
by the Snowmass 
detector.
To approximate a position
resolution of one detector element of the LHC detectors,
we introduce the resolution for the position in
$\eta$-$\phi$ of the neutral pion as a Gaussian smearing
with standard deviation $0.025/\sqrt{12}$. 
A resolution no better than a tower (as in the default Snowmass detector) was found to degrade significantly the
sensitivity to the modulation in the $\Theta$ distribution.
Finally, events are required to have at least two $\tau$-candidates
with opposite charges.  Should there be more than two
$\tau$-candidates, the highest $p_\mathrm{T}$ pair with opposite
charge is selected.  

\section{Results}
\label{s.results}
We now come to the critical question of whether the signal modulation
survives detector effects. In order to answer that question, 
consider the analysis cut-flow shown in \tref{cutflow}. 
%
\begin{table}
\begin{tabular}{|l|c|c|}
\hline\hline
Selection criterion		&		Z+j$\rightarrow \tau\tau$+j 	&	$h$+j$\rightarrow \tau\tau$+j \\
\hline\hline
None 	&		9618732					&	4698651	\\ \hline
At least three jets &								&			\\
$p_\mathrm{T} >30$\,GeV, lead jet &							&			\\
$p_\mathrm{T} >100$\,GeV	&		2445860					&	2019316	\\ \hline
Two $\tau$-candidates &							&			\\
only $\Delta$R and 	&	1069747					&	1078279	\\
isolation requirements&							&			\\ \hline
$\tau$-track $p_\mathrm{T} > 5$\,GeV&	903429					&	941368	\\ \hline
Tower $E_\mathrm{T} >  5$\,GeV	&	804566					&	856299	\\ \hline
Tower EMF $> 0.9$	&		 156140					&	160312	\\ \hline
Sum $p_\mathrm{T} > 20$\,GeV	&		  143508					&	 149591	\\ \hline
Opposite Charge &		  134628					&	 143979	\\ \hline
\end{tabular}
\caption{\label{t.cutflow}%
Cut-flows for Z$\rightarrow \tau\tau$ and $h\rightarrow \tau\tau$.
The generator and reconstruction level cuts on the jet are  $p_\mathrm{T} > 70$\,GeV, 
and 100\,GeV, respectively. See text for more details.}
\end{table}
%
The signal acceptance -- the ratio of the second number to the first
in column two of \tref{cutflow} -- is $0.430$, while that of the
background is $0.254$. The products of the efficiencies of all the
remaining cuts including the 65\% $\tau$ reconstruction
efficiency are $0.65^2 \times 143979/1078279 = 0.0564$ and $0.65^2
\times 134628/1069747 = 0.0531$  for the signal and background,
respectively.  Therefore, overall, the signal and background
efficiencies are $0.430 \times  0.0564 =  2.42 \times 10^{-2}$ and
$0.254 \times 0.0531 = 1.35 \times 10^{-2}$, respectively.

To see the effects of finite detector resolution on our ability to reconstruct the
signal $\Theta$ distribution~(\ref{e.amp2}), we present in \fref{H}
the reconstructed $\Theta$ distribution from the signal samples
generated with $\Delta=0$ before and after the inclusion of detector
effects, labelled as \emph{Monte Carlo Truth} and \emph{Delphes
Simulation}, respectively.  We perform fits to these distributions
using the functional form $c - a \cos( \gamma \Theta - 2\delta)$ with
free parameters $a$, $c$, $\gamma$, and $\delta$ in order to extract
the modulation amplitude $\alpha \equiv a / c$. The fit results are presented
in \tref{fits} and show that $\gamma\simeq1$ and $\delta\simeq0$
as expected from the analytical result~(\ref{e.amp2}) and the
assumption $\Delta = 0$, even after detector effects.  Most crucially,
the dilution of the modulation amplitude $\alpha$ due to the detector
effects is only about $\sim4\%$, going down from $0.225$ to $0.214$.

In order to check that our $\tau$ reconstruction algorithm does not
introduce artificial modulations in the background $\Theta$
distribution, we present in \fref{Z} the Monte Carlo Truth and Delphes 
Simulation $\Theta$
distributions using the background samples.  Clearly, the background is
consistent with a flat distribution after the detector effects and we
may conclude that the latter do not bias the shape of the
$\Theta$ distribution.

\begin{table}
\begin{tabular}{|c|c|c|}
\hline\hline
parameters& MC Truth & Delphes Sim.\\
\hline\hline
$c$				& $0.0397	\pm0.0001	$	& $0.0397	\pm0.0002	$\\  
$a$				& $0.0089	\pm0.0001	$	& $0.0085	\pm0.0002	$\\
$\gamma$	& $1.03	  \pm0.01		$	& $1.03		\pm0.02		$\\
$\delta$	& $-0.003	\pm0.003	$	& $0.01		\pm0.01		$\\
$\alpha \equiv a/c$	& $0.225	\pm0.002	$	& $0.214	\pm0.004	$\\
\hline
\end{tabular}
\caption{\label{t.fits}
Fits to the signal $\Theta$ distribution before and
after the inclusion of detector effects using the functional form 
$c - a \cos(\gamma\Theta-2\delta)$.}
\end{table}

\begin{figure}
\includegraphics[scale=0.45]{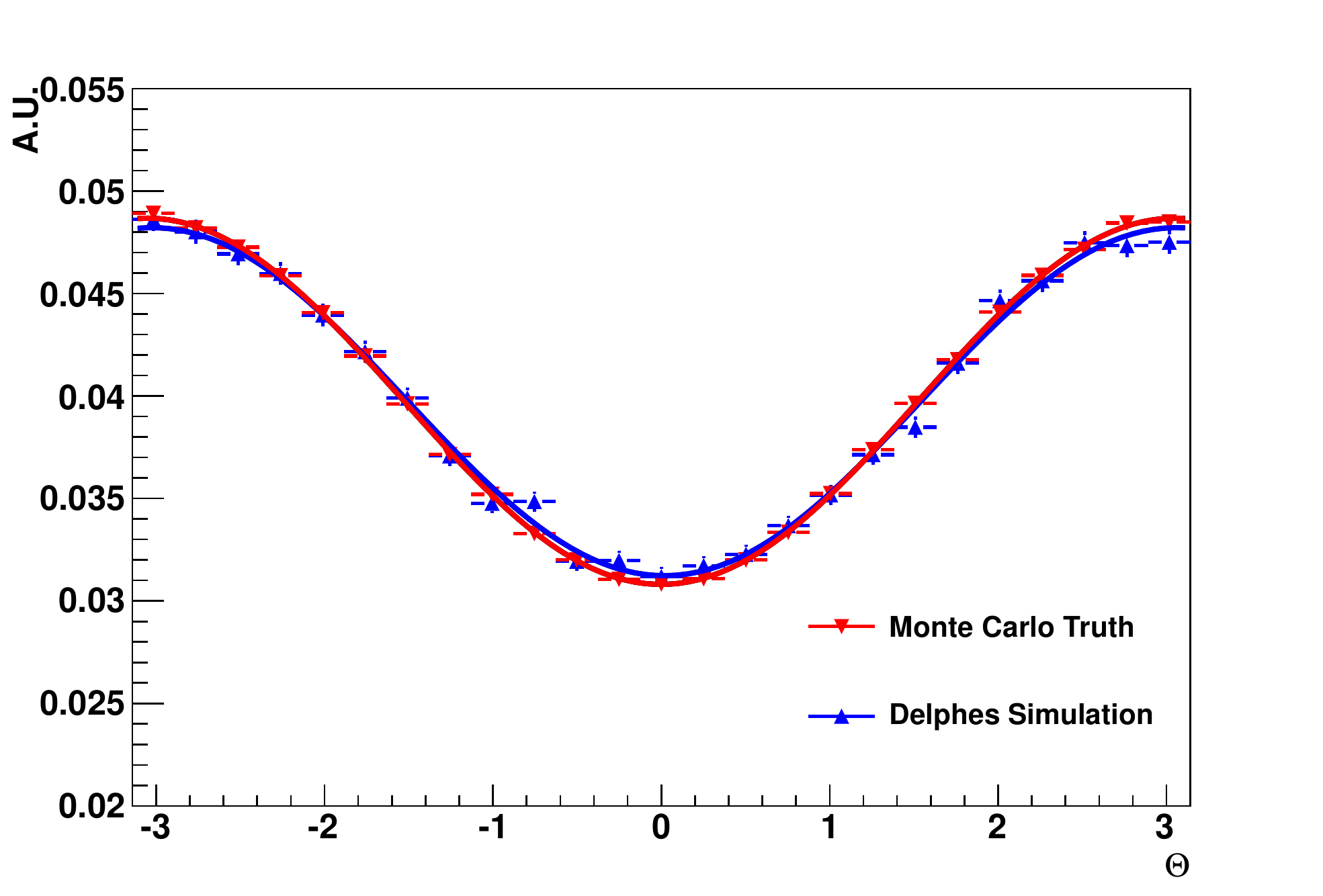}
\caption{\label{f.H}
Comparison of parton level $\Theta$
distribution (Monte Carlo Truth) with the distribution after the {\tt Delphes} simulation
and event reconstruction for $h\rightarrow\tau\tau$.}
\end{figure}

\begin{figure}
\includegraphics[scale=0.45]{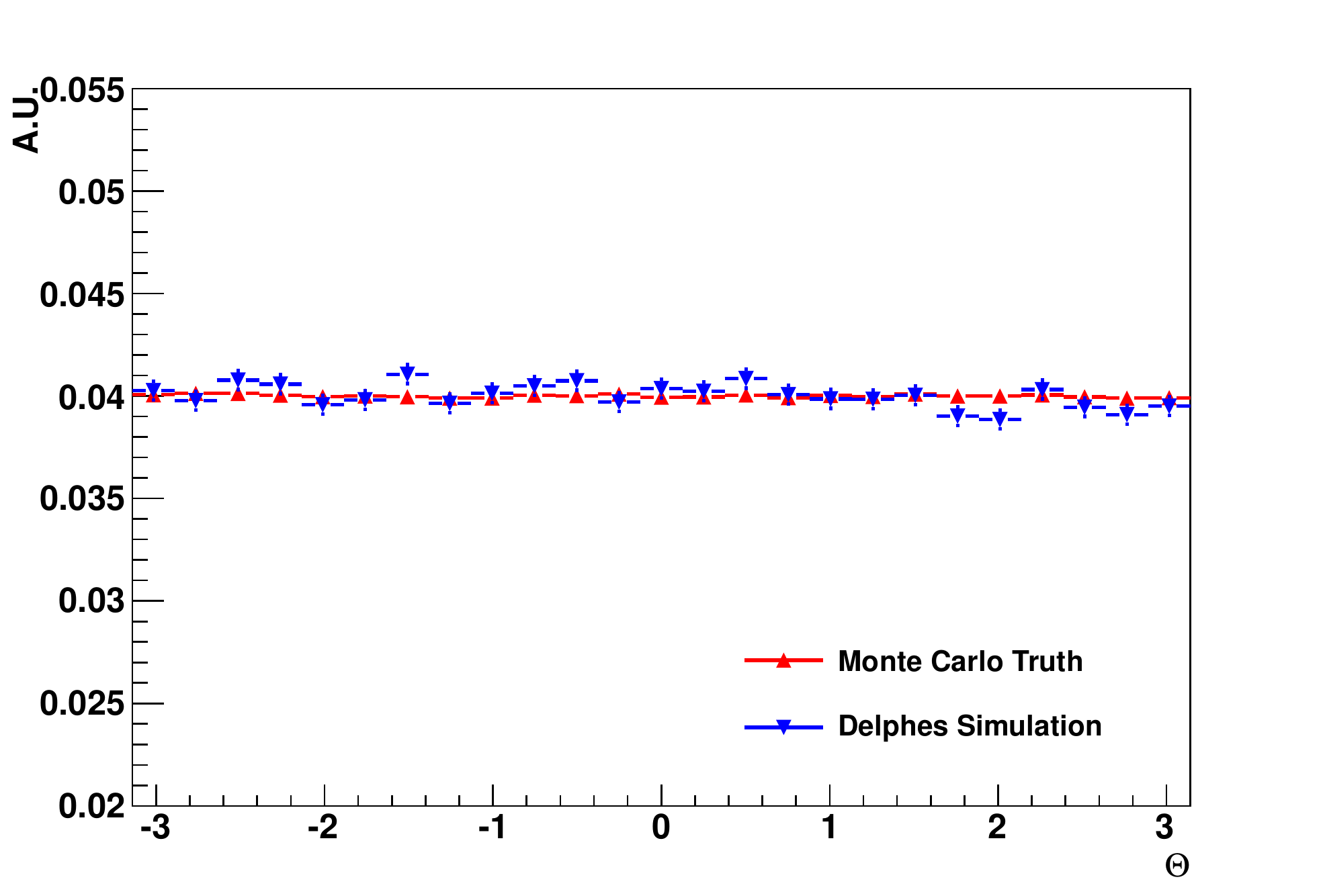}
\caption{\label{f.Z}
Comparison of parton level $\Theta$ distribution (Monte Carlo Truth)
with the distribution after the {\tt Delphes} simulation and event
reconstruction for $Z\rightarrow\tau\tau$.}
\end{figure}

Now, we address the crucial question: how well can we distinguish the
$\Theta$ distribution with $\Delta=0$ from that with $\Delta\neq0$? Given
$N$ events, each yielding a measurement of $\Theta_i$, we construct a
test statistic $t$ designed to distinguish between the SM hypothesis
$\Delta = 0$ and the alternatives $\Delta \neq 0$. A standard choice
for $t$ is
\begin{align}
t = 2 \ln \frac{L(\Delta)}{L(0)}  \,,
\label{e.t}
\end{align}
where
\begin{align}
L(\Delta)  = \prod_{i=1}^N f(\Theta_i| \Delta)
\label{e.L}
\end{align}
is the likelihood function, while $f(\Theta_i|\Delta)$ is the $\Theta$
distribution for a given value of $\Delta$. 

In order to enhance the
signal to background ratio ($S/B$), we use a boosted decision tree
(BDT), trained using the invariant mass of the reconstructed
$\tau^\pm$ system, the kinematic variables ($p_\mathrm{T},\phi,\eta$)
of the final state visible particles, and the missing transverse
energy. Therefore, each event can be characterized by the measurement
of $\Theta_i$ and the value of the BDT discriminant $d_i$,
which yields the modified likelihood function,
\begin{align}
L(\Delta)  = \prod_{i=1}^N f(\Theta_i,d_i| \Delta)  \,,
\label{e.L2}
\end{align}
where the probability density function of the data $\{(\Theta,d)\}$ is
modelled as follows,
\begin{equation}
\begin{split}
& f(\Theta,d | \Delta) 
\\
& \propto
B \, \rho^\text{B}_\text{BDT}(d)
+S \, \rho^\text{S}_\text{BDT}(d) 
\, \bigl[ 1 - \alpha \cos(\Theta - 2\Delta) \bigr]
\,,
\end{split}
\end{equation}
where $\rho^\text{S}_\text{BDT}(d)$ and
$\rho^\text{B}_\text{BDT}(d)$ are, respectively, the signal and
background BDT distributions for the discriminant parameter $d$,
while $B$ and $S$ are, respectively, the total expected background and
signal counts 
given by
\begin{equation}
B = \epsilon_\text{B} \sigma_\text{B} {\cal L}  \,,\quad
S = \epsilon_\text{S} \sigma_\text{S}  {\cal L} \,.
\label{e.BS}
\end{equation}
Here, $\sigma_\text{S}$ and $\sigma_\text{B}$ are the associated cross
section times branching fractions, the $\epsilon$'s are the signal and
background efficiencies discussed above and  ${\cal L}$ is the
integrated luminosity. 

In order to quantify how well one might be expected to distinguish 
a $\Delta \neq 0$ hypothesis from the SM hypothesis $\Delta = 0$,
we generate three samples of $\Theta$ and $d$ values 
with $\Delta = 0$, $\pi/4$, and $\pi/2$. 
For each sample, we
calculate the  values of $t$ as defined in \eref{t}, which yield the
$t$ distributions as illustrated in \fref{tdist} for the $\Delta = 0$ and $\pi/4$ samples. 
To quantify the
discrimination between the two hypothesis, the $p$-value of the SM $t$
distribution is computed using the median of the non-SM $t$
distribution. 
As proxies for systematic uncertainties in the $S$ and
$\alpha$ parameters,
the $p$-values are varied in the 10\%
neighborhood of their nominal values.  In \fref{reach} we present the
reach as a function of the integrated luminosity for
$\Delta=\pi/4$ and $\pi/2$.  Because we have chosen the ``observed" value of
$t$ to be the median of the non-SM hypothesis, the reach in this
context is interpreted as follows: if the non-SM hypothesis is true,
then there is a 50\% chance to reject the SM hypothesis $\Delta = 0$
at the $Z$-sigma level, where $Z = \Phi^{-1}(p) \equiv \sqrt2 \,
\mathrm{erf}^{-1}(2p-1)$ with $p$ being the $p$-value shown in
\fref{tdist}.

\begin{figure}
\includegraphics[scale=0.45]{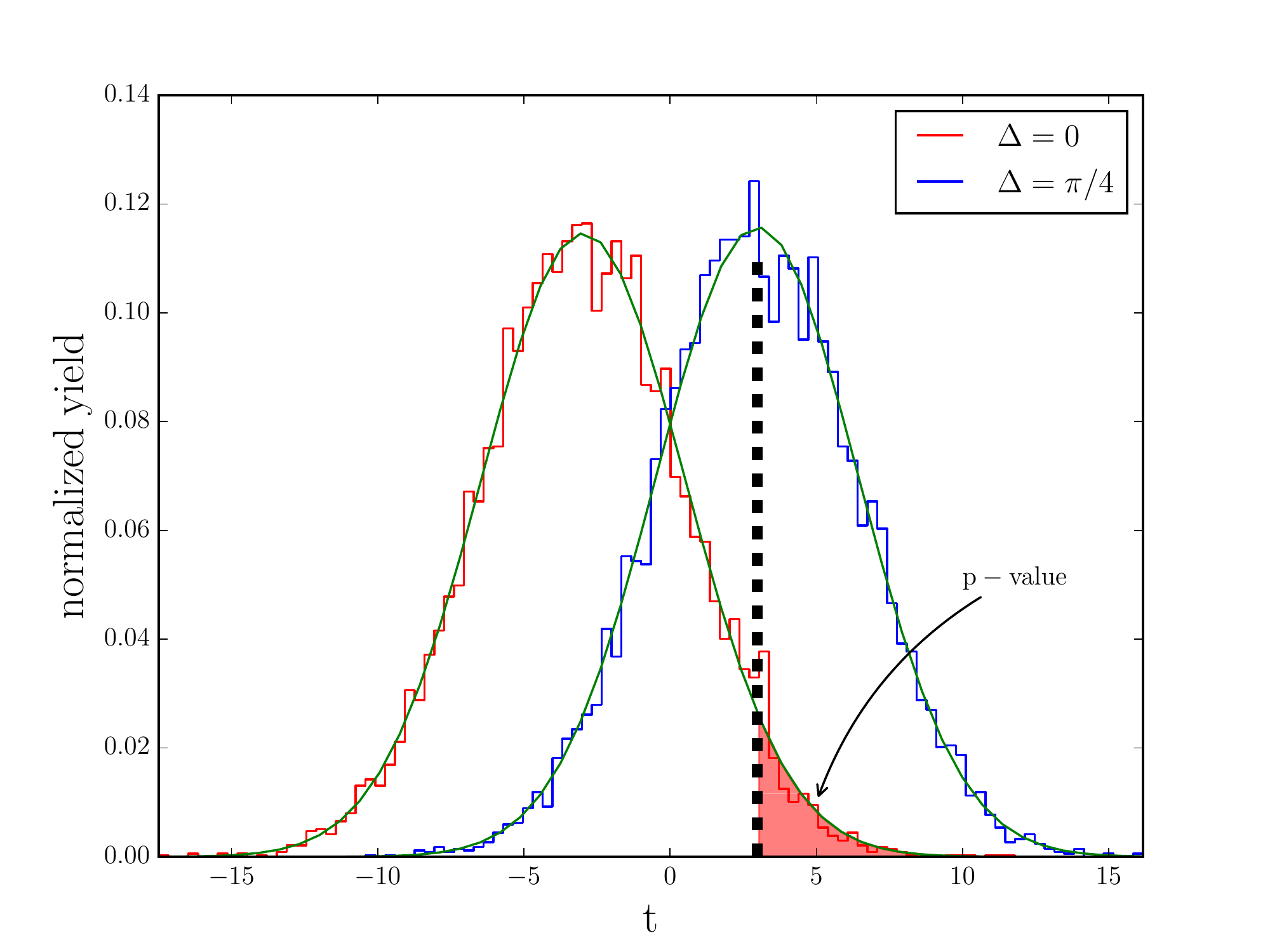}
\caption{\label{f.tdist} Distributions of the $t$ statistic for the SM
hypothesis ($\Delta = 0$) and a non-SM hypothesis ($\Delta = \pi /
4$). The ``observed" value of the statistic is taken to be the median
of the $t$ distribution of the non-SM hypothesis.}
\end{figure}

\begin{figure}
\includegraphics[scale=0.45]{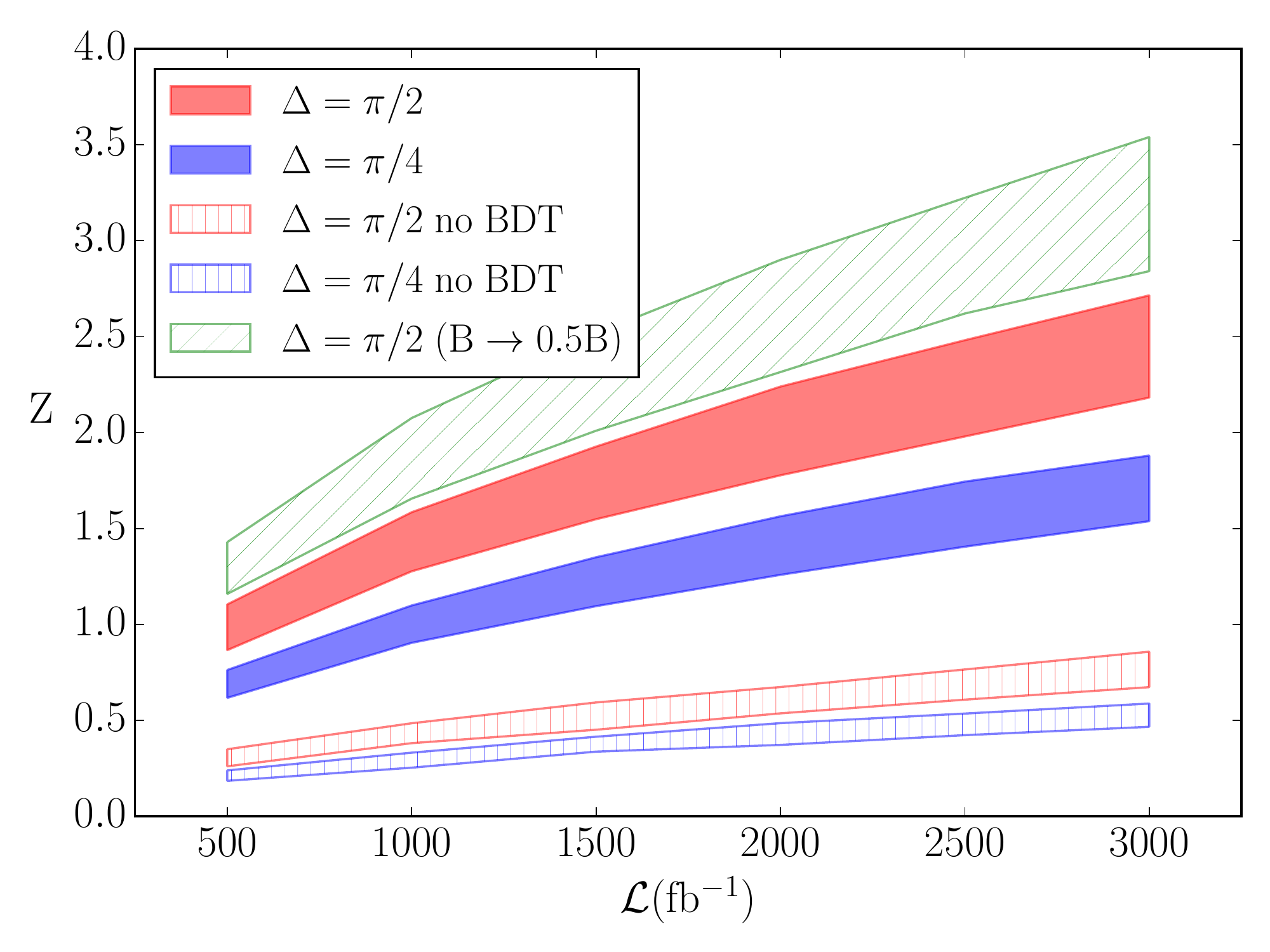}
\caption{\label{f.reach} The reach, quantified on the $Z$-sigma scale
of the normal distribution, versus the integrated luminosity. The
uppermost plot shows the effect of reducing the background by a factor
of two, while keeping every else fixed.}
\end{figure}

It is clear from \fref{reach} that the optimistic conclusion obtained
for an ideal detector does not survive the smearing effects of a
realistic detector and our reconstruction algorithms. 
At best, at $\sim 1500$ fb$^{-1}$, there is a
50:50 chance of rejecting the $\Delta = 0$ hypothesis at 95\%
confidence level. 
It is important to note where the poor performance stems from.
As mentioned earlier, the reduction of the modulation amplitude
$\alpha$ of the 
signal due to the angular resolution is negligible.
Rather, 
we find that the discrimination
power of the BDT dominantly comes from the invariant mass of the
reconstructed $\tau^+$-$\tau^-$ system, and consequently the $S/B$ ratio
is sensitive to the mass resolution, which in turn is sensitive to the
resolution of the missing transverse energy (MET)\@. 
Therefore, 
our analysis shows that the current MET resolution 
of the LHC detectors is too low to achieve a sufficient reduction of
the total background count $B$.

To illustrate the effects of improved missing transverse energy resolution, we
show in the uppermost plot in \fref{reach} the effect of increasing
$S/B$ by a factor of two by decreasing $B$ by a factor two,
while keeping everything else the same. 
Increasing the $S/B$ by
increasing $S$ by a factor two (while keeping $\alpha$ unchanged) has 
a similar but more pronounced effect.

\section{Conclusions}
\label{s.conc}
We have examined a recent proposal~\cite{Harnik:2013aja} for measuring
a potentially non-zero CP phase $\Delta$ in Higgs to $\tau^+\tau^-$
decays using an angular variable $\Theta$ that can be reconstructed
from the decay chains $\tau^\pm \to \rho^\pm \nu \to \pi^\pm \pi^0
\nu$.  
A non-zero phase of the $h$-$\tau$-$\tau$ Yukawa coupling would
be unambiguous evidence of physics beyond the SM\@.  
Contrary to previous expectations~\cite{Harnik:2013aja}, with
$\sim 1000\,\textrm{fb}^{-1}$ we find that the 
standard model CP phase, $\Delta = 0^\circ$, can be
distinguished from the $\Delta = 90^\circ$ hypothesis at no
more than 95\% confidence level.

We have addressed
two obvious effects that are expected to degrade our ability to
measure $\Delta$ from the $\Theta$ distribution.  First, we examined
the issue of neutrino momentum reconstruction, for which the original
proposal~\cite{Harnik:2013aja} employed the collinear approximation.
We have presented a theoretical argument that strongly suggests that
this cannot be improved further.

Second, we have closely studied how finite resolutions of a realistic
detector affects our ability to distinguish between the SM signal and
its deviations induced by a non-zero CP violating phase $\Delta$, using the
Snowmass detector model implemented in the {\tt Delphes} detector simulation
program.  Contrary to the naive expectation that a finite angular
resolution should hamper the reconstruction of $\Theta$ as it requires
resolving the charged and neutral pions from the $\tau$ decays, we
have found that the finite angular resolution of the LHC detectors reduces 
the modulation amplitude of the signal $\Theta$
distribution by only $\sim 4$\%, compared to that of an ideal detector. Instead, we have
found that the experimental ability to distinguish the hypotheses $\Delta=0$ from
$\Delta\neq0$ is strongly limited by the
missing transverse energy resolution.  The finite energy resolution
significantly degrades the use of the $\tau\tau$ invariant mass
distribution as the principal discriminant between the $h \to
\tau\tau$ signal and the $Z \to \tau\tau$ background.  In contrast,
the reconstruction of $\Theta$ itself is nearly unaffected by the
finite energy resolution, as the $\Theta$ variable is basically
determined by the angular topology of the Higgs boson decay products.  
Fortunately, the resolution of missing transverse energy is an
area in which improvements can be hoped for, e.g., by using
multivariate methods~\cite{Khachatryan:2014gga}.


\section*{Acknowledgment}
This work was supported in part by the US Department of Energy under
grant DE-FG02-13ER41942. The work of N. S. was partially supported by the
US Department of Energy contract No.~DE-AC05-06OR23177, under which
Jefferson Science Associates, LLC operates Jefferson Lab. 

\bibliographystyle{apsrev}


\end{document}